\documentclass[preprint2]{aastex}

\usepackage{natbib}
\usepackage{multirow}

\shorttitle{3-D reconstruction of Solar Active Regions}
\shortauthors{}

\begin{document}

\title{Reconstructing the Subsurface Three-Dimensional Magnetic Structure of A Solar Active Region Using SDO/HMI Observations}

\author{Georgios Chintzoglou and Jie Zhang}
\affil{School of Physics, Astronomy and Computational Sciences, George Mason University, 
       4400 University Dr., MSN 6A2, Fairfax, VA 22030, USA}
\email{gchintzo@gmu.edu}



\begin{abstract}
A solar active region (AR) is a three-dimensional magnetic structure formed in the convection zone, whose property is fundamentally important for determining the coronal structure and solar activity when emerged. However, our knowledge on the detailed 3-D structure prior to its emergence is rather poor, largely limited by the low cadence and sensitivity of previous instruments.   Here, using the 45-second high-cadence observations from the Helioseismic and Magnetic Imager (\emph{HMI}) onboard the Solar Dynamics Observatory (\emph{SDO}), we are able for the first time to reconstruct a 3-D datacube and infer the detailed subsurface magnetic structure of NOAA AR 11158 and to characterize its magnetic connectivity and topology. This task is accomplished with the aid of the image-stacking method and advanced 3-D visualization. We find that the AR consists of two major bipoles, or four major polarities. Each polarity in 3-D shows interesting tree-like structure, i.e. while the root of the polarity appears as a single tree-trunk-like tube, the top of the polarity has multiple branches consisting of smaller and thinner flux-tubes which connect to the branches of the opposite polarity that is similarly fragmented. The roots of the four polarities align well along a straight line, while the top branches are slightly non-coplanar. Our observations suggest that an active region, even appearing highly complicated on the surface, may originate from a simple straight flux-tube that undergoes both horizontal and vertical bifurcation processes during its rise through the convection zone. 

\end{abstract}


\section{Introduction}\label{INTRO}
The subject of AR magnetic structure in the Solar Convection Zone (SCZ, outer $\sim$ 220\,Mm of the solar radius) is one of the least understood topics but it is of crucial importance for constraining solar dynamo models and explaining what drives solar activity and space weather. It is widely believed that ARs seen on the surface are magnetic flux-tubes that are being created by the dynamo process at a depth in the SCZ (\citealt{Charbonneau_2005}). Subsequently, the flux-tubes emerge through the photospheric surface giving birth to ARs or sunspots and magnetic loop systems in the corona. On the surface, there is a high order of regularity on the pattern of AR magnetic polarities, well described by Hale's and Joy's laws (\citealt{Hale_etal_1919}). 

There is limited information on the AR three-dimensional (3-D) structure inferred from observations of AR emergence. \citet{Zwaan_1987} provided a toy-model explanation of AR emergence by attributing it to the subsurface structure of an $\Omega$-loop of magnetic flux with a frayed crest that breaks through the surface giving rise to the observed appearance of bipolar ARs. The work by \citet{Strous_et_al_1996b} and \citet{Strous_Zwaan_1999b} extends this model to include the horizontal dynamics in order to explain the fact that many flux-tubes emerge in multiple locations. \citet{Tanaka_1991} studied complex (delta configuration) ARs that exhibit rotational proper motions during their evolution (starting from Hale's-and-Joy's-laws-incompatible towards being compatible) and attributed it to a knotted 3-D topology. Also, \citet{Leka_et_al_1996} found evidence that flux-tubes emerge kink-deformed by the current they carry, thus containing a twisted 3-D structure different from an $\Omega$-shaped flux-tube. However, it is generally difficult to determine the detailed 3-D structure of an AR, due to the limitation of previous observations in terms of temporal and spatial resolution. 

On the other hand, there has been a considerable amount of theoretical work which have been developed over the past four decades trying to attack the issue computationally (for a review, see \citealt{Fan_2009}). The models of emergence in the SCZ are (a) the Thin-Flux-Tube model (TFT, \citealt{Spruit_1981}) and (b) the anelastic MHD model (\citealt{Gough_1969}) . While both models work well in the lower SCZ, they might not be valid at the top layers of the SCZ (that is, 20 - 30 Mm below surface), as the flux-tubes are not thin (TFT assumption breaks down) and the velocity field is not subsonic (anelastic approximation breaks down). The reason for making such differences lies behind the large pressure gradient at the layers close to the surface (\citealt{Fan_2009, Stein_2012}). Thus, theoretical works usually split the SCZ into two parts, the lower SCZ and the upper SCZ ($\sim$ 20\,Mm).

With the improvement of computational power and sophisticated algorithms, it has been possible for more ``realistic" numerical experiments (i.e. fully compressible, radiative-convective 3-D MHD simulations) to explore the formation of pores and sunspots (e.g. \citealt{Cheung_et_al_2010}; also see review by \citealt{Stein_2012}), although the spatial domain achieved so far is still very small (the ``deepest" simulations go down to a depth of 20\,Mm). Also, radiative-convective MHD models provide no explanation on large scale characteristics of emerging ARs $-$ such as the Joy's law of AR tilts and asymmetric foot-point separation $-$ which are reproducible by global-scale models, like the TFT approximation \citep{Caligari_et_al_1995} and anelastic MHD simulations (\citealt{Fan_2008}). To this date, because of the computational restrictions of our current era and the natural complexity of this task, there's no global, fully compressible MHD model yet capable in probing the evolution of flux-emergence throughout the entire SCZ to the surface (\citealt{Fan_2009}).

\begin{figure*}
\epsscale{2.0}
\plotone{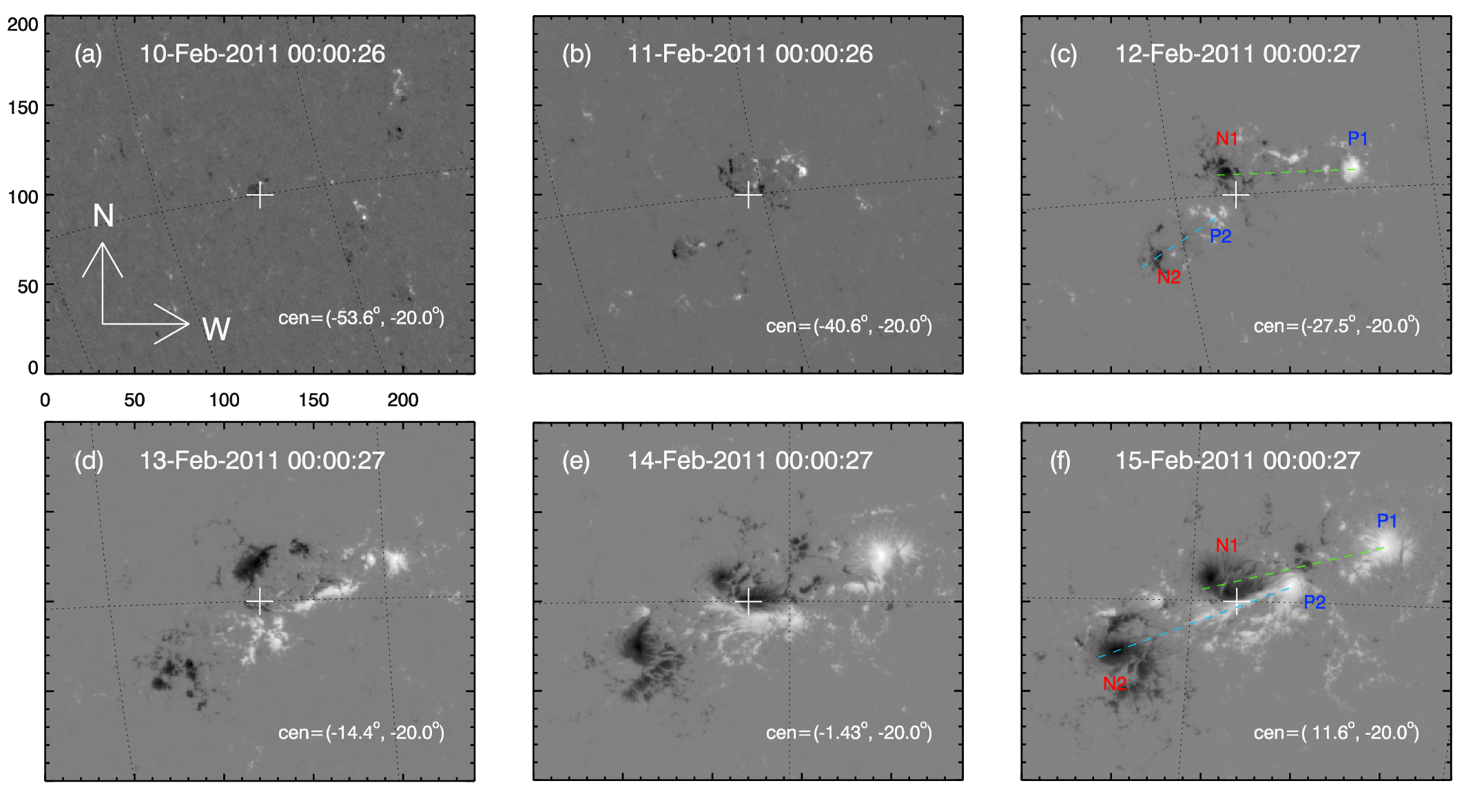}
\caption{
The first six *days of evolution of the AR 11158 as observed with the SDO/HMI LOS magnetograph. The individual polarities are named after which bipole emerged first, e.g. bipole 1, hence we name its negative polarity N1 and its positive P1, and with N2 and P2 emerging at a later time. The white cross shows the position of the guiding center of the $240 \arcsec \times 200 \arcsec$ FOV at a fixed heliographic latitude $\phi=-20^{\circ}$. Note that the bipoles initially are non-collinear (dashed lines, panel (c)); at a later time they become quasi-collinear (panel (f)).
}\label{FIG_EVOL}
\end{figure*}

In this letter, we present the implementation of an image-stacking technique as a means to reconstruct and study the 3-D structure of an emerging AR directly from observations and with great detail. The NOAA AR 11158 presented in Figure~\ref{FIG_EVOL}, is known as the one that produced the very first X-class flare (X2.2) of the Solar Cycle 24 and its energetics have been studied thoroughly by \citet{Sun_et_al_2012a}. The photospheric magnetogram images show a complex AR as seen from the surface. However, with our 3-D reconstruction method, it is rather evident that AR 11158 has a much simpler origin, i.e. the horizontal and vertical bifurcation of a single progenitor flux-tube. This is probably the first study of directly reconstructing the detailed subsurface 3-D structure and topology of solar active regions.

\section{Methodology}\label{METH} 
In this study, we used high time-cadence and high spatial resolution observations ($0\,\farcs5$/pixel) taken by the HMI instrument on board the SDO spacecraft (\citealt{Schou_et_al_2012}). The HMI instrument is able to take full-disk maps of the line-of-sight (LOS) B-field every 45 seconds. The starting time, $t_0$, of our selected observation period was on 10-Feb-2011, 00:00:28 UT when the AR 11158 first emerged at heliographic coordinates E53$\degr$S20$\degr$ and the ending time of the period under study was on 16-Feb-2011, 11:18:27 UT when the AR was at W30$\degr$S20$\degr$, well passed the central meridian. During this six-day-long period, the AR had gone through the emergence phase and fully developed into a mature region. For each of the observation frames, we performed a geometrical correction to get the radial, or normal field \emph{$B_n$}, from the LOS magnetograms. For further processing, we selected a cutout of $240\,\arcsec \times 200\,\arcsec $ with a guiding center following the AR at the Carrington rotation rate. Lastly, we rotated and remapped the solar sphere from the center of the cut-outs to the solar disk center, practically eliminating both the solar rotation and the projection effects as well, and resulting in a good alignment of cut-out images. By such pre-processing, we produced a uniform dataset ready to be used in our stacking method.

The image-stacking method works as follows. Each of the frames is a map of the B-field at the (thin) photospheric layer. We proceed onto making a stack along the time dimension using a 7.5 minute cadence of the 2-D cutouts. This cadence effectively reduces the number of images by a factor of 10, from 12\,330 images to 1233 images over the time period under study. The choice of the number of images accommodates the maximum computer capacity in both hardware (in particular, the memory) and the software.  By starting with $t_0$ at the top of the stack (the X and Y dimension) and adding images at later times consecutively at a lower height (the Z dimension),  we create a 3-D data cube, which can be used to infer the 3-D subsurface magnetic structure of the AR prior to its emergence. This technique is based on the assumption that the subsurface AR emerges as a solid body, i.e. the observed flux on the surface at each time instance corresponds to one particular height of the body. However, we know that an emerging AR is subjected to structural changes due to turbulence in the SCZ and the near-surface processes. In particular, the near-surface processes may dominate the structure of weaker field (more in the discussion section). Nevertheless, the AR selected in this paper is a strong AR with fast flux-emergence, thus making the surface effect minimal. As a first-order approximation, the velocity of the emergent structure is assumed constant, thus each frame contributes equally to the height of the structure. For better showing the emergent structure, we used only $\sim$ 4.4 days worth of data (ending on 14-Feb-2011 04:25:17 UT), i.e. the first 800 images instead of the full processed dataset. The (x, y, z) final dimensions of the datacube are 480 pix $\times$ 400 pix $\times$ 800 pix. The datacube, which is produced in IDL, is imported to PARAVIEW, a visualization software package, for further processing and inspection.

\section{Results}\label{RESUL}

\subsection{3-D Topology of the AR}
In Figure~\ref{FIG_3D}, we present the resulted reconstruction by showing the iso-surfaces in the 3-D data-cubes for two representative constant contour values (see online material for a video fly-by around the two iso-surfaces). One contour level is at 1100 G, which shows the ``skeleton" of the structure, i.e. the core structures of each sunspot/magnetic element of the AR, and  the other one at 400 G, which shows finer structures that envelope the ``skeleton" structure of 1100 G.  Apparently, the AR is composed of four major magnetic concentrations or polarities, as indicated as P1, N1, P2, N2, respectively in the bottom panel of Figure~\ref{FIG_3D}. The same four polarities are marked in panel (c) and (f) in Figure~\ref{FIG_EVOL}. As seen from the 3-D map, the four polarities originate from two bipoles, P1-N1 and P2-N2.  The positive polarity P1 connects to its corresponding negative polarity N1, and P2 connects to N2. In other words, the pairs P1-N1, P2-N2 form two neighboring flux-tube systems. It is easy to note that P1 and N1, and P2 and N2 are not closing at their apex, since at this location, the field is weak and mostly transverse to the LOS. However, if we go down to lower B-field iso-surfaces, i.e. 400 G, we can clearly see an almost closed system of adjacent branch-like arches.

Instead of coherent, solid flux-tubes, we observe a very fragmented, branch-like appearance in all polarities of the  bipoles. On the surface, such fragmentation appeared as the continuous emergence of individual small magnetic elements.  However,  these small magnetic elements exhibit a remarkably ordered, ``swarm"-like collective behavior, separating in terms of polarity and coalescing in 3-D into big ``tree-trunks" $-$ i.e. the four  polarities of the quadrupolar AR. This tree-branch-trunk feature signifies a deeper relation of all the small magnetic features within a large-scale emerging structure. 
 Each polarity  consists of several branches, which are almost perfectly connected to the branches in the opposite polarity along the flux-tubes. The branches are probably caused by a bifurcation process which is further discussed below.

\subsection{Bifurcation in Height}
\begin{figure*}
\epsscale{2.0}
\plotone{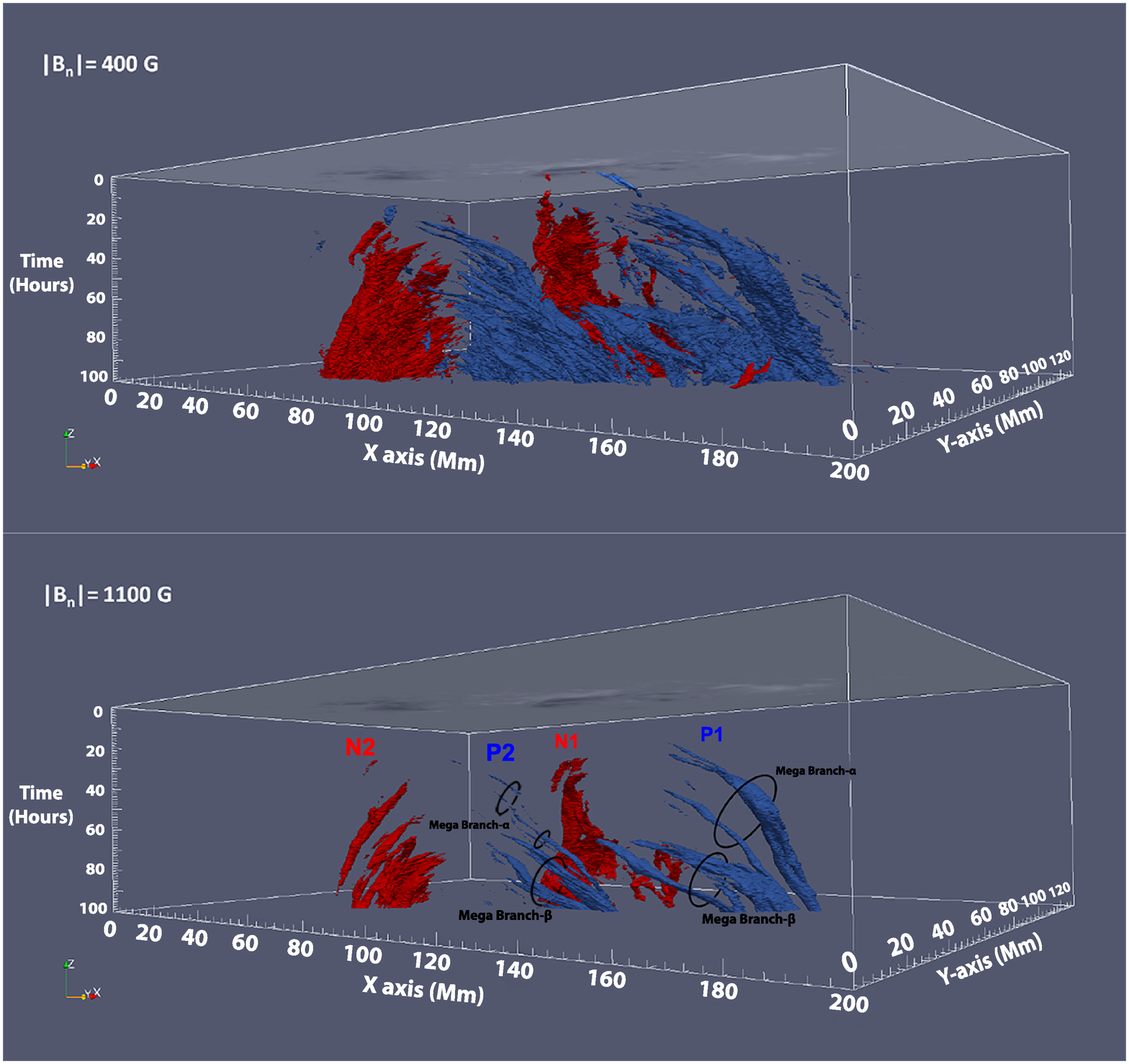}
\caption{
The 3-D reconstruction of AR 11158 using the image-stacking method on SDO/HMI LOS magnetograms. The flux-tubes shown here are at $|B_n|= 400$ G (up) and 1100 G (down). On top of each datacube is the last HMI LOS frame of the cube, i.e. the bottom frame, on 14-Feb-2011 04:25:57. The presented duration of the observation is 100.4 hours. The positive X-axis direction is westward and Y-axis northward. The length of the X-axis is roughly comparable with the height of the Solar Convection Zone (SCZ), i.e. about 200 Mm. The black loops are grouping the emergence episodes into Mega-Branches-$\alpha$,$\beta$.
}\label{FIG_3D}
\end{figure*}

Both bipoles exhibit similar bifurcation along the height (or time). An inspection on the 3-D data-cubes reveals that, for each bipole, there seems to be a two-phase evolution $- $ or, equivalently, a ``grouping" of the individual small branches to just two $-$but larger$-$ groups we dub ``Mega-Branches" (``Mega-Branch-$\alpha$" and ``Mega-Branch-$\beta$" or for short ``MB$\alpha$" and ``MB$\beta$", with ``MB$\alpha$" preceding ``MB$\beta$" in time, as illustrated in Figure~\ref{FIG_3D}). A more thorough study of the magnetic topology should include the time-flux profile to characterize the temporal evolution in a more quantitative manner as discussed below.

\begin{figure*}
\epsscale{2.0}
\plotone{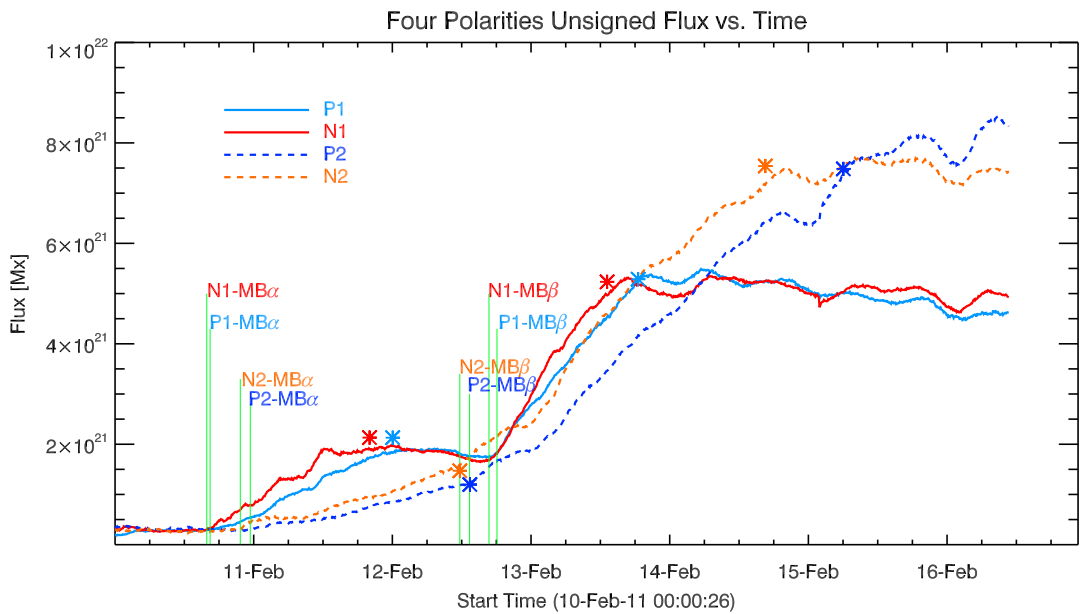}
\caption{
The time evolution of the flux for the bipoles N1-P1 (solid lines) and N2-P2 (dashed) suggests that both have an early emergence phase (i.e. ``$\alpha$") or ``front" of flux,  followed by a strong flux ``surge" (``$\beta$") as also can be seen in Fig~\ref{FIG_3D}. The Mega-Branches-$\alpha$,$\beta$ are the major contributors for each episode. Note the persistent lagging of the positive, i.e. leading polarities, P1 and P2 with respect to the following N1 and N2. The green lines denote the onset of emergence episodes. Also, the end of emergence is shown with an asterisk (*) in the respective color.
}\label{FIG_FLUX}
\end{figure*}

In Figure\,\ref{FIG_FLUX}, we show the magnetic flux-versus-time for each individual polarity of the AR, i.e. P1, N1, P2 and N2, and we overplot each polarity's unsigned profile in the same graph. According to this plot, P1-N1 (solid lines) is the first one that emerges and it is followed by the second bipole, P2-N2, about only six hours later (dashed lines). From the time-flux profile in Fig.\,\ref{FIG_FLUX}, we see that for an individual bipole (i.e. N1-P1 and N2-P2) the magnetic flux of its positive and negative polarity is $-$ to a first order $-$ very similar; such similarity is in phase throughout the emergence period. Also, for both bipoles, two major flux-emergence phases/episodes are identified, with the initial one being moderate (``$\alpha$"-Episode) and the later one (``$\beta$"-Episode) being a  stronger ``flux-surge". The major contribution to these two flux-emergence episodes is coming from the respective adjacent ``Mega-branches" of the emerging flux-tube,  as  suggested by the 3-D reconstruction of Fig.~\ref{FIG_3D}. In each of the bipoles, the Mega-Branch that arrives first at the photosphere, ``MB$\alpha$", appears somewhat weaker and fragmented whereas the branch that arrives $\sim$ 2 days later, ``MB$\beta$", is much stronger. Thus, for each bipole we have a bifurcation in height, as deduced from Figs.~\ref{FIG_3D},~\ref{FIG_FLUX}.

In Table 1, we provide the total unsigned flux emerged for each episode's polarity along with the information on the duration of emergence and the flux-emergence rate, selected manually from Figure~\ref{FIG_FLUX}. The duration of emergence is the time between the onset (annotated green lines) and the end-of-emergence (colored asterisks) for each episode. The rate of emergence is defined as the flux measured for each episode, divided by its respective duration of emergence. For comparing the individual polarities, we present measurements for $\alpha$+$\beta$-Episodes, i.e. unifying the episodes by using the onset time of $\alpha$-Episode's and for the end time, the one of $\beta$-Episode's (thus yielding information on the emergence of the individual polarities). The AR 11158 is a strong AR where emergence lasted 110 hours with a flux-emergence rate of 5.99$\times\,10^{16}$\,Mx s$^{-1}$, leading to a total emerged unsigned flux of 2.4$\times\,10^{22}$\,Mx. Going down to the level of the individual polarities, we should be able to quantify the similarities seen in the 3-D visualization. For both bipoles, the $\beta$-Episode emergence rates are a factor of 3.2 larger as compared to the $\alpha$-Episode rates, on average. Furthermore, the onset times for the Episodes of the Bipoles N1-P1/N2-P2 are only $\sim$ 6 hours apart, i.e. very similar as it can be also seen in the 3-D reconstruction of Fig~\ref{FIG_3D}.

\subsection{ Bifurcation in Horizontal Direction}
\begin{figure*}
\epsscale{2.0}
\plotone{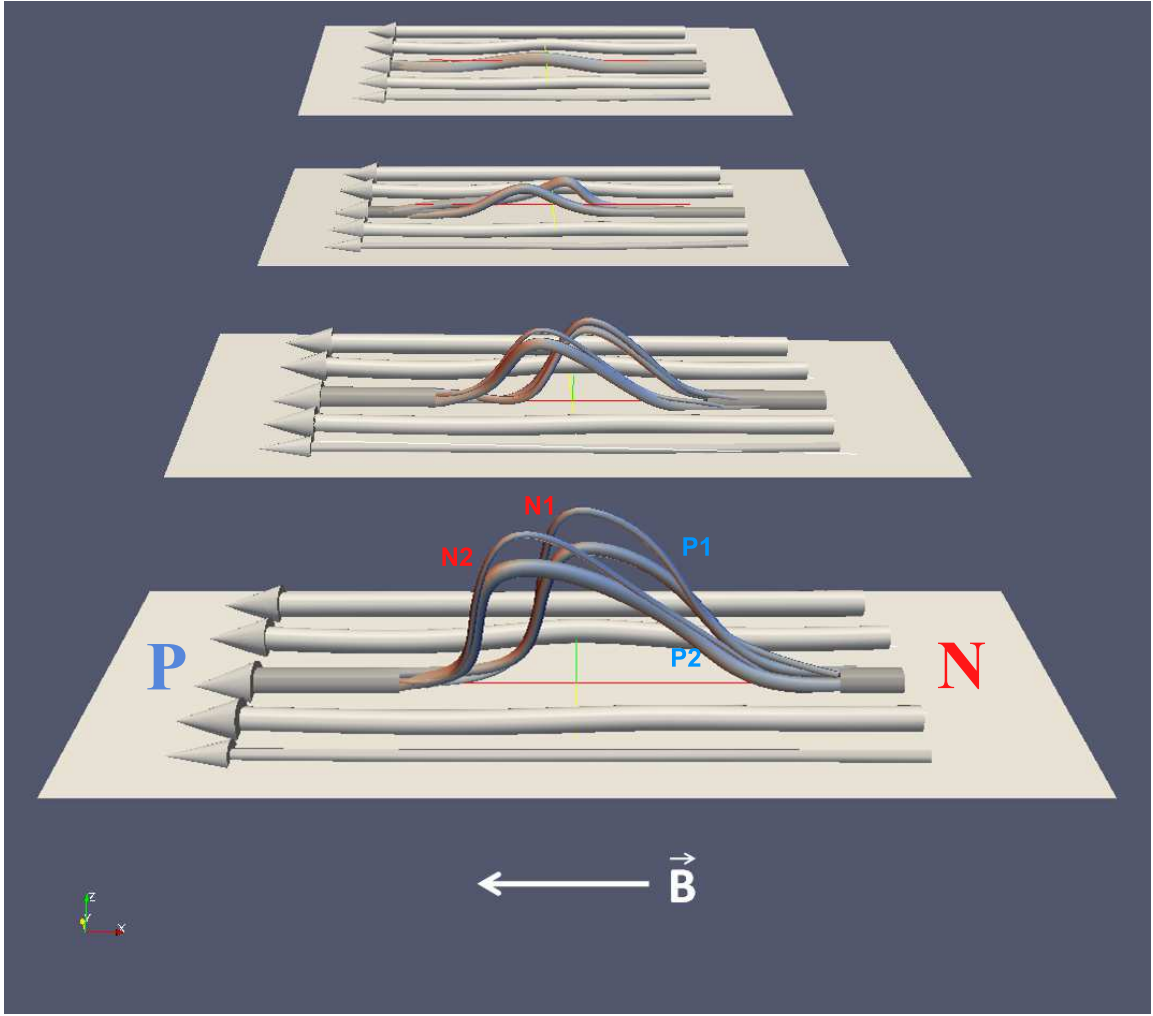}
\caption{
Model sketch of the emergence process of AR 11158 in the sub-photosphere. The plane is at the depth of the bottom of the SCZ that B-fields originally reside. The cylindrical structures shown at that plane in the SCZ are the flux-tubes of the toroidal field, generated by the solar dynamo process. For flux-tubes created in the South hemisphere during Solar Cycle 24 (like AR 11158), the B-field vector along the tubes is directed from West-to-East (here from right-to-left), as dictated by Hale's law of polarity and the Babcock-Leighton dynamo theory. The flux-tubes also develop an asymmetric lambda-shape (``\textit{$\Lambda$}") as they rise.
}\label{FIG_SKETCH}
\end{figure*}

At first sight, AR 11158's magnetic topology seems rather complex. However, the tendency for collinearity of the four polarities at the bottom of the 3-D cube, suggests that both flux-tubes might be related, even originating from the same parental magnetic flux-tube. This picture is therefore suggesting that a single sub-photospheric flux-tube has been bifurcated into two tubes along the horizontal direction , as illustrated in Figure~\ref{FIG_SKETCH}. The almost in-phase evolution of the fluxes with time of the two bipoles further reinforces this interpretation. 

Flux-tubes are non-coplanar when they first emerge through the quiet sun (see Fig~\ref{FIG_EVOL} and the online videos). However, right before the emergence of the ``flux-surge" in both of the bipoles, there is a strong pushing-aside of the already emergent Mega-Branches (``MB$\alpha$'s") in a manner like they seem to ``know in advance" about the arrival of the stronger ``flux-surge" tubes (or ``MB$\beta$'s"), suggesting that the ``MB$\beta$'s" interact/collide sub-photospherically with the trunks of the ``MB$\alpha$" tubes. At a later time, i.e. at the bottom of the data-cube, the polarities assume the ``correct", Joy's law-compliant tilt.

Further, for each bipole, the asymmetric polarity separation suggests flux-tubes with an oblique $\Lambda$-shape instead of axisymmetric $\Omega$-loops. The $\Lambda$-shape has its leading leg more tilted away from the vertical direction than the trailing leg.  This asymmetry can be understood in terms of the Coriolis force acting on a rising flux-tube, as discussed by \citet{Caligari_et_al_1995} with the TFT approximation and also reproduced using anelastic MHD models (\citealt{Abbett_et_al_2001}).

\section{Discussion and Conclusion} 

In this letter, we presented a novel image-stacking technique for reconstructing  the 3-D structure of buoyant flux-tubes rising through the solar surface and forming observed solar active regions.  Sequences of images have been used before to infer  the structure of ARs , e.g. \citet{Tanaka_1991} and \citet{Leka_et_al_1996}. However, the previous attempts are limited to tracking the locations of AR centroids with time, but not the entire structure.  To our best knowledge,  this work is probably the first true implementation of the image-stacking technique to reconstruct the detailed 3-D structure of an AR, using advanced visualization software and high-cadence high-resolution magnetogram data.  

At least in the early stages of emergence, the emerging magnetic structures are two non-coplanar neighboring bipoles, but a more detailed picture reveals a bifurcated structure for both bipoles, in the horizontal direction and along the height as well. In the low B-field iso-surfaces, multiple magnetic arches can be observed to emerge in a continuous manner, with the like-polarities coalescing  with time. The 3-D reconstruction provided  good evidence that Mega-branches could be originating from the same flux-tube below the photosphere. Last, we find that there's a dual-phase evolution for both bipoles, as suggested by both the topology in 3-D  and the  time-flux profile of the AR, providing further evidence for a bifurcation in height. Observations also indicate that the two bipoles have a common origin. The two bipoles have a similar topology in 3-D, similar temporal evolution in flux-emergence, and most significantly, appear almost collinear at the later stage of emergence. It is possible that the two bipoles are the result of bifurcation of a single progenitor flux-tube early in the evolution.

It is interesting to note that the  3-D topology of the AR 11158 $-$for each bipole as well as overall$-$ exhibits all the qualitative characteristics of the TFT approximation. The TFT model's successes in reproducing observations qualitatively are well known (\citealt{Caligari_et_al_1995}). The same qualitative characteristics of the TFT were reproduced in incompressible MHD simulations by \citet{Abbett_et_al_2001} by including the Coriolis force due to the solar rotation, in order to study its effects on the fragmentation of flux-tubes. This simulation reproduces the non-axisymmetrical topology that arises due to the Coriolis force. From our reconstruction it is evident that we have such an asymmetry. The fact that we observe it suggests that the upper SCZ has not a severe impact in altering the magnetic topology of flux-tubes while traversing the lower SCZ, after they are born. However, the small arch-like magnetic ``fibers'' seen in the 400\,G isosurfaces may be caused by the strong surface processes due to a large pressure gradient in the upper SCZ.

This study also demonstrates that the image-stacking technique is a promising method for studying the 3-D structure of ARs prior to their emergence. In the future, we will study the magnetic vector 3-D structure by fully using the magnetic vector observations from the SDO/HMI instrument.


\acknowledgments
The authors wish to thank Drs. M. Linton,  Y. Fan and M.K. Georgoulis for valuable discussions. G. Chintzoglou also thanks  Prof. C.E. Alissandrakis for inspiring conversation and encouragement. We acknowledge the support from NSF ATM-0748003, NSF AGS-1156120.  
One of the authors (G.C.) was supported by NASA Headquarters under the NASA Earth and Space Science Fellowship Program - Grant NNX12AL73H.

%

\bibliography{/home/gchintzo/paper/references}

\clearpage

\clearpage

\clearpage

\clearpage

\clearpage

\begin{deluxetable}{cccccrrc}
\tablecolumns{8}
\tablewidth{0pc}
\tablecaption{Flux-Emergence Rate, Total Flux Emerged and Duration of Emergence}

\tablehead{
\colhead{Bipole} & \colhead{Polarity} &\colhead{Episode\tablenotemark{*}} &\colhead{Rate\tablenotemark{a}}  &\colhead{Total\tablenotemark{b}}   & \colhead{Start Time}   & \colhead{End Time} &\colhead{Duration\tablenotemark{c}} } 
\startdata
          &    & $\alpha$           &  1.33 & 1.52   &  10-Feb 16:27   & 12-Feb 00:00\phn & 31.66\phn \\ 
\cline{3-8}
          & P1 & $\beta$            &  3.87 & 3.41   &  12-Feb 18:06   & 13-Feb 18:30\phn & 24.50\phn \\ 
\cline{3-8}
    1     &    & $\alpha$+$\beta$ &  1.85 & 4.93   &  10-Feb 16:27   & 13-Feb 18:30\phn & 74.15\phn \\
\cline{2-8}
          &    & $\alpha$           &  1.57 & 1.59   &  10-Feb 15:52   & 11-Feb 20:03\phn & 28.16\phn\\
\cline{3-8}
          & N1 & $\beta$            &  4.43 & 3.27   &  12-Feb 16:31   & 13-Feb 13:10\phn & 20.50\phn\\
\cline{3-8}
          &    & $\alpha$+$\beta$ &  1.87 & 4.66   &  10-Feb 15:52   & 13-Feb 13:10\phn & 69.28\phn \\
\cline{1-8}
          &    & $\alpha$           &  0.67 & 0.91   &  10-Feb 23:24   & 12-Feb 13:21\phn & 37.99\phn\\
\cline{3-8}
          & P2 & $\beta$            &  2.64 & 6.15   &  12-Feb 13:21   & 15-Feb 06:00\phn & 64.64\phn \\          
\cline{3-8}
    2     &    & $\alpha$+$\beta$ &  1.91 & 7.06   &  10-Feb 23:24   & 15-Feb 06:00\phn & 102.63\phn \\
\cline{2-8}
	  &    & $\alpha$           &  0.91 & 1.24   &  10-Feb 21:40   & 12-Feb 11:40\phn & 38.00\phn\\ 
\cline{3-8}
          & N2 & $\beta$            &  2.94 & 5.59   &  12-Feb 11:40   & 14-Feb 16:30\phn & 52.87\phn\\
\cline{3-8}
          &    & $\alpha$+$\beta$ &  2.09 & 6.83   &  10-Feb 21:40   & 14-Feb 16:30\phn & 90.88\phn \\
\cline{1-8}
1+2       &    &                      &  5.99 &23.79   &  10-Feb 15:52   & 15-Feb 06:00\phn & 110.16\phn \\

\enddata
\tablenotetext{*}{The Mega-Branches-$\alpha$,$\beta$ are the major contributors to the corresponding emergence episodes.}
\tablenotetext{a}{ $\times 10^{16}$\,Mx\,s$^{-1}$}
\tablenotetext{b}{ $\times 10^{21}$ Mx}
\tablenotetext{c}{ hours}
\label{TABLE_Stat}
\end{deluxetable}

\end{document}